# Evaluating a Signalized Intersection Performance Using Unmanned Aerial Data[*]

Mujahid I. Ashqer[1], Huthaifa I. Ashqar[2], Mohammed Elhenawy[2], Mohammed Almannaa[3], Mohammad A. Aljamal[4], Hesham A. Rakha[5], *IEEE, Fellow*, and Marwan Bikdash[6]

*Abstract*—This paper presents a novel method to compute various measures of effectiveness (MOEs) at a signalized intersection using vehicle trajectory data collected by flying drones. MOEs are key parameters in determining the quality of service at signalized intersections. Specifically, this study investigates the use of drone raw data at a busy three-way signalized intersection in Athens, Greece, and builds on the open data initiative of the *pNEUMA* experiment. Using a microscopic approach and shockwave analysis on data extracted from realtime videos, we estimated the maximum queue length, whether, when, and where a spillback occurred, vehicle stops, vehicle travel time and delay, crash rates, fuel consumption, $CO_2$ emissions, and fundamental diagrams. Results of the various MOEs were found to be promising, which confirms that the use of traffic data collected by drones has many applications. We also demonstrate that estimating MOEs in real-time is achievable using drone data. Such models have the ability to track individual vehicle movements within street networks and thus allow the modeler to consider any traffic conditions, ranging from highly under-saturated to highly over-saturated conditions. These microscopic models have the advantage of capturing the impact of transient vehicle behavior on various MOEs.

## I. Introduction

Signalized intersections are major elements in causing abrupt changes in traffic patterns that lead to high traffic delays, fuel consumption, and emissions. However, traffic signals are crucial in increasing traffic safety along the intersection by avoiding conflicts of various movements. Traffic engineers carefully design traffic signal controllers to efficiently operate traffic movements with the aim of minimizing traffic queues, delays, fuel consumption, and emissions. Traffic signal controllers rely on information that captures traffic demands for various movements along the intersection. This information can be fed from different detection techniques, such as the traditional inductive loop detectors and camera systems. Recently, researchers have developed traffic signal controllers using advanced detection techniques such as probe vehicles (vehicles with GPS/Bluetooth devices and connected vehicles). These advanced techniques allow traffic signal controllers to use real-time information that assists in improving traffic stream efficiency.

This study utilizes row data collected using a new detection technique, namely, Unmanned Aerial Vehicles (UAVs) (i.e., drones) at a busy signalized intersection in Athens, Greece. The study investigates the use of UAVs data to determine measures of effectiveness (MOEs) for signalized intersections. The UAVs technology has shown a promising future in a wide range of applications such as transportation, security, monitoring, survey, environmental mapping, etc. Furthermore, UAVs technology is easy to deploy, introduces minimum interruptions to the studied area's traffic stream, and is relatively low-cost to maintain.

Traffic delays and queues are principal measures for determining the capacity and the traffic quality of service of signalized intersections. Calculating the number of vehicles and the queue length helps estimate the performance measures of an intersection. Many studies have used several methods to estimate the elements of MOEs, including using the Kalman Filtering technique to estimate the number of vehicles traveling along signalized approaches using real-time probe vehicle data [1, 2] and applying Lighthill–Whitham–Richards (LWR) shockwave theory [3]. Liu et al. used LWR shockwave theory to identify traffic state changes that distinguish queue discharge flow from upstream arrival traffic. Moreover, Ban et al. proposed methods to estimate real-time queue lengths at signalized intersections using sample travel times from mobile traffic sensors. The estimation was based on the observation that critical pattern changes of intersection travel times or delays. The model and algorithm were tested using field experiments and simulation data [4].

[1] Ph.D. candidate at Computational Data Science and Engineering at North Carolina AT State University, USA. (email: miashqer@ncat.edu).
[2]Senior Transportation Engineer at Precision Systems, Inc. and Adjunct Professor at University of Maryland Baltimore County, USA. (email: hiashqar@vt.edu).
[2] Research Fellow at CARRS-Q, Queensland University of Technology, Australia. (email: mohammed.elhenawy@qut.edu.au).
[3] Assistant Professor at Civil Engineering Department, King Saud University, Saudi Arabia. (email: malmannaa@ksu.edu.sa).
[4] Manager at Grant Thornton LLC, USA. (email: m7md92@vt.edu).
[5] Samuel Reynolds Pritchard Professor of Engineering, Charles E. Via, Jr. Dept. of Civil and Environmental Engineering and the Bradley Department of Electrical and Computer Engineering (courtesy), Director of the Center of Sustainable Mobility, VTTI, Virginia Tech. (email: hrakha@vt.edu).
[6] Professor and Chair at Department of Computational Data Science and Engineering at North Carolina AT State University, USA. (email: bikdash@ncat.edu).

Traffic delays and queues are key parameters in determining the quality of service of signalized intersections. In this study, we used a microscopic approach and shockwave analysis on data extracted from a real-time video, which were collected using a drone, to characterize the various MOEs on a signalized approach for a three-way signalized intersection in Athens, Greece. Specifically, for the specified area in green shown in Fig. 1, we used Unmanned Aerial Data to estimate the measures of effectiveness, including the maximum queue length, whether, when, and where a spillback occurred, vehicle stops vehicle travel time, and delay, crash rates, and fundamental diagrams.

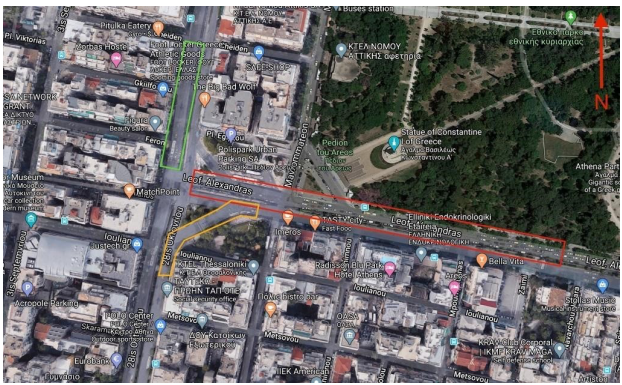

Fig. 1: Study area.

## II. RELATED WORK

Numerous studies were conducted on UAV video data to analyze and improve transportation networks. Recent research efforts utilized UAV datasets to detect queue and estimate its length on freeways and intersections [5, 6], identify and track heavy vehicles [7], investigate crash risk [8], detect hazard obstacles and accidents [9, 10], and better monitor traffic violations [11]. In the area of estimating queue lengths using UAV data, there are limited efforts conducted at signalized intersections. Freenmn et al. employed UAV to capture the traffic formulation at two signalized intersections in Kuwait [12]. They first utilized two drones to capture the entire area around the intersection while it was fully congested. Two periods with different traffic volumes were used. The authors were able to estimate the stationary stacking headway for individual vehicles, and types of vehicles were estimated. However, this study only used static UAV video and measured the stacking gab graphically at a stationary status.

Khan et al. has used UAV video data to develop analytical methodology at a four-way intersections in Sint-Truiden, Belgium [7]. The authors extracted the individual trajectories of vehicles and drew the flow fundamental diagrams, dividing them into three states. Afterward, they conducted a shockwave analysis macroscopically and calculated the queue length for each direction based on the flow fundamental diagrams and shockwave analysis. Yet, the authors did not investigate and consider the microscopic level to calculate the queue length and other MOEs. Ke et al. adopted Machine Learning (ML) techniques such as K-means and Kanadelucas-Tomasi algorithms to estimate both macroscopic and microscopic traffic parameters [13]. Although they have achieved a high accuracy, the computational time remains an obstacle for ML algorithms, especially when it comes to estimating the real-time queue length. In a different research effort, Ke et al. proposed a framework to estimate seven traffic flow fundamental parameters for both macroscopic and microscopic levels, yet queue length and other MOEs were not investigated [14].

Unlike the previous related works, this study proposes a microscopic methodology to estimate the MOEs of an approach in a signalized intersection using UAV data. Specifically, our approach can be used dynamically for real-time estimation of MOEs using UAV data, which could be used for traffic management applications. Our approach uses a microscopic method based on shockwave analysis, which considers the mobility and interaction of individual vehicles and deemed to be more accurate than macroscopic methods. While our approach is believed to achieve high accuracy in estimating MOEs, it effectively accomplishes that with a relatively less computational complexity and time.

## III. DATASET

This study used a 14-min video, which was recorded by a UAV at a three-way signalized intersection in Athens, Greece, as shown in Fig. 1 and Fig. 2. The video was filmed in sunny weather by a drone hanging over a sufficiently high point at the center of the intersection to cover three approaches as follows: 1) Leof. Alexandras Road, with direction towards the west to 28is Oktovriou Road (Red polygon with 300m length), 2) 28is Oktovriou Road, with direction towards north to Leof. Alexandras Road (Yellow polygon with 100m length), and 3) 28is Oktovriou Road, with direction towards south to 28is Oktovriou Road (Green polygon with 100m length). The geometric design of the intersection consists of five lanes (including a left-turn pocket) for the green polygon, three lanes for the red one, and two lanes for the orange one. The red polygon contains two additional traffic signals, and the speed limit for the three polygons is 55 km/h (34 mph).

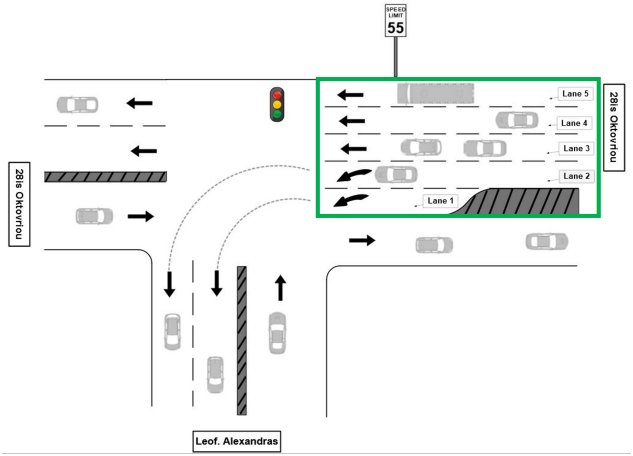

Fig. 2: Intersection layout.

The video was then converted to second-by-second trajectories for all types of vehicles, including cars, taxis, powered two-Wheeler's, buses, and medium and heavy vehicles as shown in Fig. 3. Each trajectory has a unique tracking ID with vehicle type in addition to initial traveled distance and average speed (once they enter the polygon). Each point in the trajectory has a timestamp (in seconds), instantaneous latitude and longitude, instantaneous vehicle speed, instantaneous latitude, and longitude acceleration/deceleration. These features were utilized to identify the starting of each point of the trajectory and determine the lane allocations for each vehicle. More information on the methodology will be provided in the following two sections.

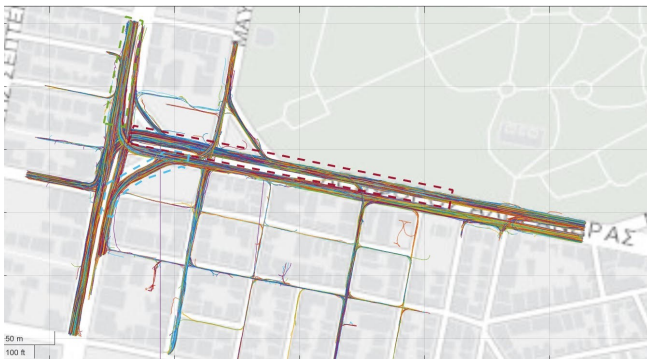

Fig. 3: Illustration of extracted trajectories.

The first step in using the UAV data entailed utilizing Google Earth to divide each polygon into smaller subpolygons representing the lanes in the green area shown in Fig. 1 and Fig. 2. We used a Keyhole Markup Language (KML) file, which is an international standard maintained by the Open Geospatial Consortium, Inc. (OGC), to display the geographic data from Google Earth. KML files were created by pinpointing the resulted locations of adding an image overlay of the study area. This geographic data was used in identifying the lane (i.e., sub-polygon) that the vehicles were on during their trip within the green-defined area. Keeping in mind that many vehicles may change their lane multiple times during their trip, the instantaneous longitude and latitude of each vehicle. That said, in order to validate the lane allocations at the upstream edge of the defined area and after leaving the downstream, the origin-destination of each trajectory was identified, and the mismatched values were corrected. Subsequently, a lane-specific time-space diagram was created using vehicle trajectories that were associated with the determined sub-polygon (i.e., lane). The third step entailed applying shockwave analysis on the microscopic level to determine the backward formation and backward recovery waves for each lane-specific queue. This was used to mainly identify the queues and the spillbacks past the upstream edge of the polygon, which is considered the zero reference of our study. The results of this process and other information were used in estimating the other MOEs for the defined area on the signalized intersection, as will be described in Section IV. It is worth mentioning here that we assumed that a vehicle traveling at a speed less than or equal to the typical pedestrian speed of 4.5 km/h (1.2 m/s) is stopped.

IV. PROBLEM FORMULATION AND ESTIMATION APPROACHES

*A. Problem Formulation*

The new era of sharing information and "big data" has raised expectations to make mobility more predictable and controllable through a better utilization of data and existing resources. The realization of these opportunities requires going beyond the existing traditional ways of collecting traffic data that are based either on fixed-location sensors or GPS devices with low spatial coverage or penetration rates and significant measurement errors, especially in congested urban areas [15]. *pNEUMA* is a first-of-its-kind experiment aiming to create the most complete urban dataset to study congestion. A swarm of 10 drones hovering over the central business district of Athens, Greece over multiple days to record traffic streams in a congested area of a 1.3 km$^2$ area with more than 100 km-lanes of road network, around 100 busy intersections (signalized or not), many bus stops and close to half a million trajectories. The aim of the experiment is to record traffic streams in a multi-modal congested environment over an urban setting using UAV that can allow the deep investigation of critical traffic phenomena. The *pNEUMA* experiment develops a prototype system that offers immense opportunities for researchers. This open science initiative creates a unique observatory of traffic congestion, a scale an-order-of-magnitude higher than what was available till now, that researchers from different disciplines around the globe can use to develop and test their own models [15].

This study is built on the open data initiative of *pNEUMA*, which is a unique dataset that was acquired during a firstof-its-kind experiment using a swarm of drones over a dense city center of Athens. This dataset consists of more than half a million detailed trajectories of almost every vehicle that was present in the study area. The dataset includes trajectories that were monitored by one single drone covering a wide area over the central district of Athens, Greece. Both major and minor roads, bus stops and signalized intersections are included in the study area. The dataset includes trajectories from cars, taxis, powered two-wheelers, buses, medium and heavy vehicles [15]. In this study, we used part of *pNEUMA* dataset to estimate the measures of effectiveness including the maximum queue length, whether, when and where a spillback occurred, vehicle stops, vehicle travel time and delay, crash rates, and extract the fundamental diagrams.

As mentioned in Section III, the first phase of our proposed approach was dividing each polygon into mini polygons, representing the lanes into each polygon using Google Earth. We visually determined the mini polygons (i.e. lanes) for each direction. The second phase was identifying which lane the vehicles were on. Using the instantaneous longitude and latitude, we were able to track each vehicle from upstream to downstream the specified area. Then, we created timespace diagrams (vehicle trajectories) and then linked this to the associated lane for each polygon. Fig. 4 depicts a spacetime diagram resulted from this process. The third phase was using shockwave analysis to determine when the queue starts spilling back at the onset of the yellow/red signal traffic indication. This information along with instantaneous vehicle counts helped us characterize the signal phase timing for the traffic light at the intersection. For verification purposes, we matched between the vehicle counts per lane with the OD matrix (calculated from the trajectories). This process is summarized in Fig. 5.

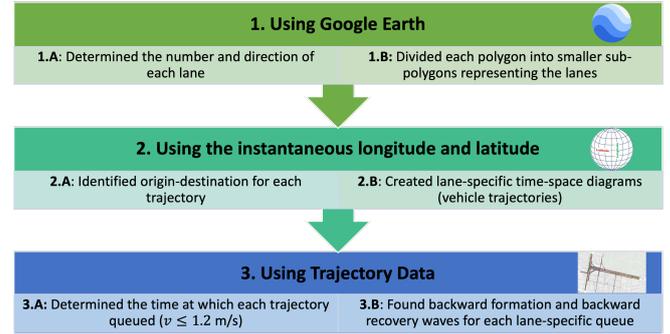

Fig. 5: Summary of proposed approach.

### B. Computation of Queue Information

We assumed that a vehicle is queued when its speed is less than or equal to the typical pedestrian speed of 1.2 m/s. We constructed the time-space diagram for each lane in the approach. For each vehicle in the stream, we moved along its trajectory from the furthest upstream record within the defined area to its defined area exit record (see the green area shown in Fig. 1 and Fig. 2. We then identified the time stamps and locations (relative to the defined area upstream point) of the points at which the vehicles enter (backward forming shockwave) and exit (backward recovery shockwave) a queuing state. The furthest upstream record that the trajectory enters a queuing state was located. The spatial extent of a queue is computed relative to the downstream end of the defined area, which equals to the area length minus the distance relative to the furthest upstream area boundary. Spillbacks were identified in the area when entry into queuing was immediately prior to the upstream edge of the area.

### C. Estimation of Travel Time

The model we developed determines the travel time for any given vehicle by providing that vehicle with a *time card* upon its entry to any of the identified links. Subsequently, this time card is retrieved when the vehicle leaves the link. The difference between these entry and exit times provides a direct measure of the link travel time experienced by each vehicle as Equation (1) and Equation (2) show.

$$N(t) = N(t - \Delta t) + u(t) \qquad (1)$$

$$TT(t) = H(t) \times N(t) \qquad (2)$$

where $N(t)$ is the number of vehicles traversing the link at time $t$, $N(t - \Delta t)$ is the number of vehicles traversing the link in the previous time interval, and $u(t)$ is the system inputs, as described in Equation (1). Equation (2) represents the system output by measuring the average travel time for the CVs. $H(t)$ is a transition vector that converts the vehicle counts to travel times, as shown in Equation (2).

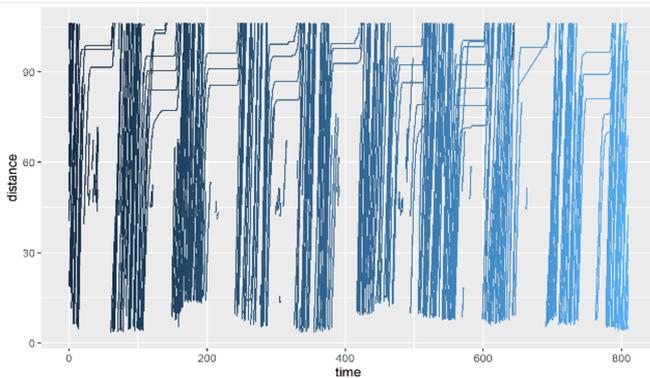

Fig. 4: Space-time diagram.

### D. Estimation of Vehicles Stops

Each time a vehicle decelerates, the drop in speed is recorded as a partial stop, as demonstrated in Equation (3) [16]. The sum of these partial stops is also recorded. This sum, in turn, provides a very accurate explicit estimate of the total number of stops that were encountered along that particular link. This means that this method will often report that a vehicle has experienced more than one complete stop along a link. Multiple stops arise from the fact that a vehicle may have to stop several times before ultimately clearing the link stop line. This finding, while seldom recorded by or even permitted within macroscopic models, is a common observation within actual field data for links on which considerable over-saturation queues exist.

$$S(t_i) = \frac{u(t_i) - u(t_{i-1})}{u_f} \quad (3)$$

where;

$S(t_i)$: Instantaneous partial stop estimate at time $t_i$
$u(t_i)$: Speed at instant $t_i$ $t_{i-1}$: Speed at instant $t_{i-1}$ $u_f$: Roadway free-flow speed

### E. Estimation of Vehicle Delay

The model estimates vehicle delay every deci-second as the difference in travel time between travel at the vehicle's instantaneous speed and travel at free-speed, as indicated in Equation (4) [17]. This model has been validated against analytical time-dependent queuing models, shockwave analysis, and the Canadian Capacity Guide, Highway Capacity Manual, and Australian Capacity Guide procedures [17, 18].

$$d(t_i) = \Delta t(1 - \frac{u(t_i)}{u_f}) \quad (4)$$

where:

$\Delta t$ : The data processing time step (0.1 seconds in our case).

### F. Estimation of Crash Rates

The safety model that was used here is based on US national crash statistics. The model computes the crash risk for 14 different crash types as a function of the facility speed limit and a time-dependent measure of exposure. The use of a time-dependent measure of exposure allows the model to capture differences in the crash risk that result from differences in the network efficiency. The model also computes the vehicle damage and level of injury to the passengers involved in the crash based on the vehicle's instantaneous speed. The use of the instantaneous speed means that the crash damage and injury level is responsive to the level of congestion. Consequently, the model can capture the safety impacts of operational-level alternatives including Intelligent Transportation Systems. By multiplying the distance-based crash rate by the facility free-speed, it was possible to estimate a time-based crash rate. The advantage of a timebased crash rate is that the rate level of exposure increases with higher levels of congestion even though vehicles might not necessarily travel longer distances. More information can be found here [19].

$$CrashRate_i = e^{a_1^i * u_f + a_2^i} \quad \forall 1:15 \quad (5)$$

Where:

$a_1$: Regression Coefficient $a_1$. $a_2$: Regression Coefficient $a_2$. $u_f$: Free-flow Speed

### G. Fuel Consumption

To find Fuel consumption, we used the Virginia Tech Comprehensive Power-Based Fuel Consumption Model (VTCPFM). VT-CPFM is a microscopic fuel consumption model based on instantaneous power; the detailed VT-CPFM model is described in the original paper by Rakha et al., which also includes a MATLAB script to run the model [20]. Other models either require calibration of specific parameters from laboratory or field testing or produce a bang-bang control. VT-CPFM avoids both of these problems: because data collection is not always feasible, VT-CPFM uses only publicly available data. Additionally, because the function for fuel consumption is a second degree polynomial with respect to vehicle specific power (VSP), the partial derivative with respect to torque is a function of torque and the bang-bang control is not produced [20]. The model also can be used for different vehicle classes including light-duty vehicles [20], heavy-duty vehicles [21], and buses [22].

First, power is calculated using Equation (6), as follows:

$$P(t_i) = \left(\frac{R(t_i) + 1.04ma(t_i)}{3,600\eta_d}\right) * v(t_i) \quad (6)$$

where:

$P(t_i)$ = power at time step $t_i$ (kW), $m$ = vehicle mass (kg), $a(t_i)$ = vehicle acceleration at time step $t_i$ (m/s²), $v(t_i)$ = vehicle speed at time step $t_i$ (km/h), $\eta_d$ = driveline efficiency, and

$R(t_i)$ = resistance force at time step $t_i$ (N).

The resistance force is calculated with Equation (7), as follows:

$$R(t_i) = \frac{\rho}{25.92}C_D C_h A_f v(t_i)^2 + 9.8066m\frac{C_r}{1,000}(c_1 v(t_i) + c_2)$$
$$+ 9.8066mG(t_i) \quad (7)$$

where:

$\rho$ = density of air (1.2256 kg/m³ at sea level and 15°C), $C_d$ = vehicle drag coefficient (unitless),

$C_h$ = correction factor for elevation [which equals 1 0.085$H$ where $H$ is elevation (km)], $A_f$ = vehicle frontal area (m²), $G(t_i)$ = roadway grade at time step $t_i$, and $c_r$, $c_1$, and $c_2$ = rolling resistance parameters (unitless) [20].

Then, fuel consumption, $FC(L/s)$, is calculated by using Equation (8). The alpha values are calculated by using time, power, and fuel consumed from the EPA city and highway test cycles. A detailed list of required variables and potential sources for the VT-CPFM can be found in [20].

$$FC(t_i) = \begin{cases} \alpha_0 + \alpha_1 P(t_i) + \alpha_2 P(t_i)^2 & \forall P(t_i) \geq 0 \\ \alpha_0 & \forall P(t_i) < 0 \end{cases} \quad (8)$$

where:

$\alpha_0$, $\alpha_1$ and $\alpha_2$ are vehicle-specific model constants that are calibrated for each vehicle.

### H. Fundamental Diagrams

To construct the fundamental diagrams using UAV data in this study, we used a steady-state car-following model, namely, Van Aerde car-following model, which was proposed by Van Aerde [23] and Van Aerde and Rakha [24], which combines the Pipes and Greenshields models into a singleregime model. The model, which requires four input parameters, can be calibrated using field loop detector data. The details of the calibrating procedure are beyond the scope of this paper but can be found here [25]. The functional form of the Van Aerde model amalgamates the Greenshields and Pipes car-following models, as demonstrated in Equation 9. This combination provides the functional form with an additional degree of freedom by allowing the speed-atcapacity to be different from the free-speed. Specifically, the first two parameters provide the linear increase in the vehicle speed as a function of the distance headway, while the third parameter introduces curvature to the model and ensures that the vehicle speed does not exceed the free-speed. The addition of the third term allows the model to operate with a speed-at-capacity that does not necessarily equal the free-speed, as is the case with the Pipes model.

In summary, the Van Aerde single-regime model combines the Greenshields and Pipes models in order to address the main flaws of these models. Specifically, the model overcomes the shortcoming of the Pipes model in which it assumes that vehicle speeds are insensitive to traffic density in the uncongested regime, which has been demonstrated to be inconsistent with a variety of field data from different facility types [26]. Alternatively, the model overcomes the main shortcoming of the Greenshields model, which assumes that the speed-flow relationship is parabolic, which again is inconsistent with field data from a variety of facility types [26]. It is sufficient to note at this time that the calibration of the car-following model requires estimating four parameters, namely $c_1$, $c_2$, $c_3$ and $k$ utilizing Equations 10, 11, 12, and 13. These four parameters are a function of the roadway free-speed, the speed-at-capacity, capacity, and jam density. The detailed derivation of these four equations is presented in [25, 26] using a number of boundary conditions including the maximum flow and jam density boundary conditions. One aim of this study is to use a field data extracted from drones to calibrate Van Aerde car-following model and construct the corresponding fundamental diagrams.

$$h = c_1 + c_3 u + \frac{c_2}{u_f - u} \quad (9)$$

$$m = \frac{2u_c - u_f}{(u_f - u_c)^2} \quad (10)$$

$$c_2 = \frac{1}{k_j \left(m + \frac{1}{u_f}\right)} \quad (11)$$

$$c_1 = mc_2 \quad (12)$$

$$c_3 = \frac{-c_1 + \frac{u_c}{q_c} - \frac{c_2}{u_f - u_c}}{u_c} \quad (13)$$

where:

$c_1$ = fixed distance headway constant (km), $c_2$ = first variable distance headway constant (km²/h), $c_3$ = second variable distance headway constant (h), $u_f$ = free-speed (km/h), $u_c$ = speed at capacity (km/h), $q_c$ = flow at capacity (veh/h), $k_j$ = jam density (veh/km), and $m$ = constant used to solve the three headway constants (h/km).

### V. RESULTS AND DISCUSSION

Using a microscopic approach and based on shockwave analysis that considers the mobility and interaction of individual vehicles, we estimated the MOEs of an approach to a signalized intersection using UAV data. We constructed the time-space diagrams for each lane on the studied approach, as shown in Fig. 1 and Fig. 2. The green polygon consists of five lanes (four lanes plus a left turn pocket lane). We labeled the pocket lane as follows: the leftmost lane as Lane 1 and increased the counter until the rightmost lane, which is labeled Lane 5. The total number of vehicles in the area is 750 during the monitoring period. We also identified the distribution of the vehicle types in each lane as shown in TABLE I. In the table, a vehicle may be counted multiple times as we were able to capture the vehicles that make lane-change movement from a lane to another during the monitoring period (See TABLE II). However, a vehicle is counted one time in each lane at the end. We also investigated the 95$^{th}$ percentile and found it to be about 42 km/hr as shown in Fig. 6.

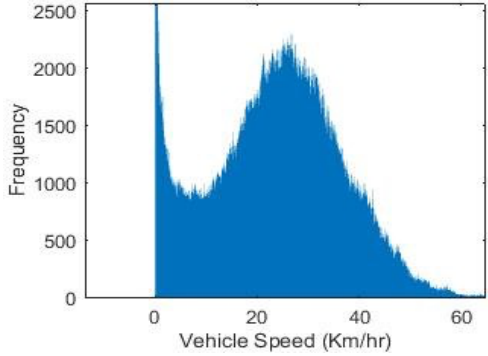

Fig. 6: Vehicle speed distribution.

TABLE I: Total number of vehicles for each Vehicle types per lane.

| Vehicle Type | Lane 1 | Lane 2 | Lane 3 | Lane 4 | Lane 5 |
|---|---|---|---|---|---|
| Light- & medium-duty | 124 | 210 | 277 | 195 | 69 |
| Motorcycle | 24 | 53 | 124 | 160 | 66 |
| Heavy-duty | 1 | 3 | 4 | 4 | 3 |
| Bus | 0 | 1 | 1 | 9 | 11 |
| Total | 149 | 267 | 406 | 368 | 149 |

TABLE II: Lane changes of each Vehicle type.

| Vehicle Type | Number of Vehicles | Number of Lane Changes |
|---|---|---|
| Light- & medium-duty | 491 | 384 |
| Motorcycles | 237 | 190 |
| Heavy-duty | 8 | 7 |
| Bus | 14 | 8 |
| Total | 750 | 589 |

### A. Queue Information

Using the information extracted from the time-space diagram of each lane, we were able to identify the length of each queue that was formed in each lane during the monitoring period, as shown in TABLE III. The table shows the queue length that occurred on each lane, when, and where it was occurred (i.e. the latitude (lat) and longitude (long) of the beginning of the queue and the end). The table demonstrates that the maximum queue length in the green polygon occurred in Lane 2 (the leftmost upstream of the left turn pocket lane) at a length of 102.7m (between the coordinates of (37.99225, 23.73141) and (37.99280, 23.73154)), which happened at time of 350.2s after the beginning of the monitoring period. Moreover, we used the time-space diagram to identify spillbacks on each lane during the monitoring period. TABLE IV summarizes the results for the various spillbacks. It shows that there were two spillbacks as follows: one in Lane 2 (middle-left lane), which is also recorded as the maximum queue length; and the other one in Lane 3 (middle lane) - both occurred at time 350.2s after the beginning of the monitoring period.

TABLE IV: The spillbacks information.

| Lane | Timestamp (s) |
|---|---|
| Lane 2 | 350.20 |
| Lane 3 | 350.20 |

### B. Travel Time

Our model to determine the travel time on each lane is a microscopic based on Equation (1) and Equation (2). The model determines the link travel time for any given vehicle by providing that vehicle with a time card upon its entry and exit from any link. The results of estimating the travel time on each lane per the vehicle types is shown in TABLE V and per the traffic movement of each vehicle type is shown in TABLE VI. The tables show that the maximum travel time occurred by a heavy-duty vehicle on Lane 3 (i.e., part of the through traffic movement) and is equal to about 32s. They also show that Lane 5 has the highest average travel time for all vehicles traveling through the study area. This could be as it is the closest lane to the parking lane and to the curbside, which pedestrians usually use. The second highest average travel time was for Lane 2, which is part of the left turn traffic movement.

TABLE V: Travel time on each lane per the number of vehicle types (s).

| Vehicle Type | Lane 1 | Lane 2 | Lane 3 | Lane 4 | Lane 5 |
|---|---|---|---|---|---|
| Light- & medium-duty | 20.29 | 23.27 | 19.48 | 13.58 | 13.14 |
| Motorcycles | 9.02 | 9.70 | 10.53 | 9.93 | 30.20 |
| Heavy-duty | 2.08 | 4.80 | 31.99 | 22.74 | 9.28 |
| Bus | - | 1.40 | 5.36 | 9.45 | 16.09 |
| Total | 18.35 | 20.29 | 16.84 | 11.99 | 20.84 |

TABLE VI: Travel Time per movement for each vehicle type (s).

| Vehicle Type | Left Turn (Lane 1 & 2) | Through (Lane 3, 4, & 5) |
|---|---|---|
| Light- & medium-duty | 22.17 | 16.55 |
| Motorcycles | 9.49 | 13.97 |
| Heavy-duty | 4.12 | 22.43 |
| Bus | 1.40 | 12.73 |
| Total | 19.60 | 15.55 |

### C. Vehicle Stops

To calculate vehicle stops, we used a microscopic model that computes instantaneous partial and full stops for undersaturated and oversaturated conditions in signalized intersection by using second-by-second speed measurements as described in in Equation (4) [17]. This model have shown that there is a significant impact of vehicle stops on fuel consumption and emissions [27]. The model indicates that the vehicle fuel consumption rate is more sensitive to cruise-speed levels than to vehicle stops [27]. TABLE VII shows the number of vehicles stops on each lane per the number of vehicle types. It shows that vehicles on Lane 2 experienced the highest number of stops, followed by Lane 3. It also shows that heavy-duty vehicles experienced the highest number of stops, which occurred in Lane 3, followed by light- and medium-duty vehicles experience in Lane 2. Moreover, TABLE VIII shows that the total number of vehicles turning left experienced stops more than vehicles going through the intersection.

TABLE III: The queue information on each lane.

| Lane | Queue length (m) | Timestamp (s) | Start lat | Start long | End Lat | End long |
|---|---|---|---|---|---|---|
| Lane 1 | 25.0 | 565.04 | 37.99190 | 23.73136 | 37.99211 | 23.73140 |
| Lane 2 | 102.7 | 350.20 | 37.99225 | 23.73141 | 37.99280 | 23.73154 |
| Lane 3 | 102.2 | 346.00 | 37.99229 | 23.73139 | 37.99280 | 23.73150 |
| Lane 4 | 97.6 | 529.56 | 37.99275 | 23.73148 | 37.99275 | 23.73148 |
| Lane 5 | 98.6 | 783.12 | 37.99200 | 23.73121 | 37.99277 | 23.73142 |

TABLE VII: Vehicles stops on each lane per the number of vehicle types.

| Vehicle Types | Lane 1 | Lane 2 | Lane 3 | Lane 4 | Lane 5 |
|---|---|---|---|---|---|
| Light- & medium-duty | 0.12 | 0.40 | 0.28 | 0.23 | 0.19 |
| Motorcycle | 0.11 | 0.21 | 0.24 | 0.21 | 0.24 |
| Heavy-duty vehicle | 0.09 | 0.11 | 0.68 | 0.26 | 0.04 |
| Bus | - | 0.01 | 0.07 | 0.10 | 0.13 |
| Total | 0.12 | 0.36 | 0.27 | 0.20 | 0.20 |

TABLE VIII: Vehicle stops per movement for each vehicle type.

| Vehicle Type | Left Turn (Lane 1 & 2) | Through (Lane 3, 4, & 5) |
|---|---|---|
| Light- & medium-duty | 0.30 | 0.24 |
| Motorcycle | 0.18 | 0.22 |
| Heavy-duty vehicle | 0.10 | 0.35 |
| Bus | 0.01 | 0.11 |
| Total | 0.27 | 0.23 |

*D. Vehicle Delay*

Delay at signalized intersections is considered a significant factor in determining the level of service at the intersection approaches as well as a parameter that is used in the optimization of traffic signal timings. In this study, we used a microscopic approach to obtain delay estimates in the study area. Within the model, delay is estimated for each individual vehicle by calculating, for each traveled link, the difference between the vehicle's recorded travel time and the travel time that the vehicle would have experienced on the link at free speed, as shown in Equation (4) [17]. TABLE IX shows the results of travel delay on each lane of the study area divided by the number of each vehicle type. When all vehicle types are considered, Lane 2 has the highest delay of all lanes, followed by Lane 5 and Lane 1. This depicts that the vehicles turning to left are more likely to experience higher delay than vehicles traveling through the intersection, which is clearly shown in TABLE X. Moreover, the highest delay recorded was for heavy-duty vehicles traveling through Lane 3 and is equal to 20.63s. TABLE X also shows that light- & medium-duty vehicles and motorcycles have higher travel delay when turning to the left. However, heavy-duty vehicles and buses have higher travel delays when traveling through the intersection.

TABLE IX: Average total delay on each lane per vehicle type (s).

| Vehicle Type | Lane 1 | Lane 2 | Lane 3 | Lane 4 | Lane 5 |
|---|---|---|---|---|---|
| Light- & medium-duty | 11.63 | 15.77 | 10.30 | 7.51 | 8.98 |
| Motorcycle | 8.55 | 7.40 | 5.57 | 4.01 | 16.68 |
| Heavy-duty vehicle | 1.18 | 1.87 | 20.63 | 13.89 | 1.53 |
| Bus | - | 0.71 | 2.75 | 3.88 | 8.12 |
| Total | 11.06 | 13.90 | 8.94 | 5.97 | 12.18 |

TABLE X: Travel delay per movement for each vehicle type (s).

| Vehicle Type | Left Turn (Lane 1 & 2) | Through (Lane 3, 4 & 5) |
|---|---|---|
| Light- & medium-duty | 14.23 | 9.12 |
| Motorcycles | 7.76 | 6.95 |
| Heavy-duty | 1.70 | 12.97 |
| Bus | 0.71 | 6.05 |
| Total | 12.88 | 8.28 |

*E. Crash Rates*

We used a safety model that is based on US national crash statistics as described in [19]. The model computes the crash risk for 14 different crash types as a function of the facility speed limit and a time-dependent exposure measure as shown in Equation 5. The model can capture the safety impacts of operational-level alternatives including Intelligent Transportation Systems (ITS). TABLE XI shows the crash rates for the 14 crash types included in the model. Results show that the rear-end crashes in the same traffic way and same direction has the highest crash rate in the intersection, followed by forward impact of a single driver. The total crash rate was found to be about 0.038 crashes for every vehicle miles traveled (VMT).

TABLE XI: Crash rates for 14 crash types (crashes/VMT).

| Crash Type | Crash Rate |
|---|---|
| Single Driver - Right Roadside Departure | 0.00296 |
| Single Driver - Left Roadside Departure | 0.00243 |
| Single Driver - Forward Impact | 0.00544 |
| Same Traffic Way and Same Direction - Rear-End | 0.00597 |
| Same Traffic Way and Same Direction - Forward Impact | 0.00023 |
| Same Traffic Way and Same Direction - Sideswipe/Angle | 0.00179 |
| Same Traffic Way and Opposite Direction - Head-On | 0.00041 |
| Same Traffic Way and Opposite Direction - Forward Impact | 0.00090 |
| Same Traffic Way and Opposite Direction - Sideswipe/Angle | 0.00244 |
| Change Traffic Way and Vehicle Turning – Turn Across Path | 0.00366 |
| Change Traffic Way and Vehicle Turning – Turn Input Path | 0.00434 |
| Intersecting Paths – Perpendicular Crash | 0.00267 |
| Backing Vehicle | 0.00043 |
| Other or Unknown | 0.00367 |
| Total Crash Rate | 0.03760 |

*F. Fuel Consumption*

Using traffic data from drones, we investigated the use of Virginia Tech Comprehensive Power-Based Fuel Consumption Model (VT-CPFM). We investigated the applicability of VT-CPFM as a microscopic fuel consumption model based on instantaneous power [20]. VT-CPFM function for fuel consumption is a second degree polynomial with respect to vehicle specific power (VSP) using Equation (6) [20]. We used VT-CPFM in this study for the different vehicle classes including light-duty vehicles [20], heavy-duty vehicles [21], and buses [22] that passed through the study area. We found that the total fuel consumption resulted from all the 750 vehicles passing through the approach is about 207L during the 14-min study period; about 0.28L per vehicle; or 0.25 L/s.

*G. Fundamental Diagrams*

We attempted to calibrate the microscopic single-regime traffic stream model and its corresponding four parameters that was proposed in Van Aerde [23] and Van Aerde and Rakha [24] using the trajectory data from drones. We used this model because while the model requires four parameters for calibration, it still provides more degrees of freedom to reflect different traffic behavior across different roadway facilities. The key to calibrate the model would be to have coverage over the full range of the fundamental diagram. Using this data, it was difficult to calibrate the fundamental diagrams given that the data lacks the sufficient coverage. Still, using datasets that cover longer periods at the same approach might be able to validate the ability of the Van Aerde model to reflect traffic stream behavior at the intersection.

## VI. CONCLUSION

Traffic delays, fuel consumption, stopping, spillbacks, and queues are key parameters in determining the quality of service at signalized intersections. This study investigates the use of drone raw data at a busy signalized intersection in Athens, Greece to find the corresponding MOEs. Collecting traffic data using flying drones is an emerged technology that has shown promising future in a wide range of applications because drones are easy to deploy, introduce minimum interruptions to the traffic stream of the studied area, and are relatively low cost to maintain.

This study builds on the open data initiative of the *pNEUMA* experiment that used a swarm of drones over a dense city center of Athens. Using a microscopic approach and shockwave analysis on the data extracted from real-time video, we determined the various MOEs of the signalized approach on a busy three-way signalized intersection in Athens, Greece. Specifically, we estimated the measures of effectiveness including the maximum queue length, whether, when and where a spillback occurred, vehicle stops, vehicle travel time and delay, crash rates, vehicle fuel consumption and emissions of $CO_2$, and fundamental diagrams. Results of the various MOEs were found to be promising, which confirms that the use of traffic data collected by drones has many advantages.

We also found that the use of microscopic approaches to computing the MOEs at signalized intersections in realtime is achievable and desirable. Such models have the ability to track individual vehicle movements within street networks and thus capture traffic conditions on various MOEs, ranging from highly under-saturated to highly oversaturated conditions. The results of these models have higher accuracy than other aggregate models given that the approach captures vehicle transient behavior. Specifically, at the delay level, microscopic models can determine the delay incurred by an individual vehicle while traveling without the need for macroscopic formulas. This allows them to evaluate uniform and overflow delay, or delays in under-saturated and over-saturated traffic conditions, allowing for the evaluation of complex traffic situations. In addition, the ability to record vehicle speed and position on a second-by-second basis further allows the recording of speed profiles and the direct estimation of deceleration, stopped and acceleration delays, crash risk, and vehicle fuel consumption levels, as demonstrated in this study.

## VII. ACKNOWLEDGEMENT

Authors would like to thank pNEUMA project for providing the dataset that was used in this study.